\documentclass{emulateapj}

\usepackage{amsmath}
\usepackage{epstopdf}
\usepackage{natbib}
\usepackage{color}
\usepackage{hyperref}
\usepackage{psfig}
\usepackage{times}
\usepackage{subfigure}
\usepackage{hyperref}
\usepackage{floatrow}

\newcommand{\be}{\begin{equation}}
\newcommand{\ee}{\end{equation}}

\newcommand{\vv}{0.21}
\newcommand{\vvv}{0.215}
\newcommand{\vvw}{0.2}
\newcommand{\ww}{0.9}

\slugcomment{Accepted to AJ}

\shorttitle{RATIR LIGO Followup}
\shortauthors{Golkhou et al.}

\def\lessim{\mathrel{\hbox{\rlap{\hbox{\lower4pt\hbox{$\sim$}}}\hbox{$<$}}}}
\def\gtrsim{\mathrel{\hbox{\rlap{\hbox{\lower4pt\hbox{$\sim$}}}\hbox{$>$}}}}

\begin{document}

\setlength{\pdfpageheight}{\paperheight}
\setlength{\pdfpagewidth}{\paperwidth}


\title{RATIR Followup of LIGO/Virgo Gravitational Wave Events} 

\author{V. Zach Golkhou\altaffilmark{1,2,3,10,11}
, Nathaniel R. Butler\altaffilmark{3,4}
, Robert Strausbaugh\altaffilmark{4,5}
, Eleonora Troja\altaffilmark{6,8}
, Alexander Kutyrev\altaffilmark{6}
, William H. Lee\altaffilmark{7}
, Carlos G. Rom\'an-Z\'u\~niga\altaffilmark{9}
, Alan M. Watson\altaffilmark{7}
}

\altaffiltext{1}{DiRAC Institute, 
Department of Astronomy, University of Washington, 
3910 15th Avenue NE, Seattle,
WA, 98195, USA}
\altaffiltext{2}{The eScience Institute, 
University of Washington, Seattle, WA 98195, USA}
\altaffiltext{3}{School of Earth and Space Exploration,
Arizona State University, Tempe, AZ 85287, USA}
\altaffiltext{4}{Cosmology Initiative,
Arizona State University, Tempe, AZ 85287, USA}
\altaffiltext{5}{Physics Department,
Arizona State University, Tempe, AZ 85287, USA}
\altaffiltext{6}{NASA, Goddard Space Flight Center, Greenbelt, MD 20771, USA}
\altaffiltext{7}{Instituto de Astronom\'ia, Universidad Nacional Aut\'onoma de M\'exico, Apartado Postal 70-264, 04510 M\'exico, D. F., M\'exico}
\altaffiltext{8}{Department of Astronomy, University of Maryland, College Park, MD 20742, USA}
\altaffiltext{9}{Instituto de Astronom\'ia, Universidad Nacional Aut\'onoma de M\'exico, Unidad Acad\'emica en Ensenada, Ensenada BC 22860, M\'exico}
\altaffiltext{10}{Moore-Sloan Fellow}
\altaffiltext{11}{Dirac Fellow}

\begin{abstract}
Recently we have witnessed the first multi-messenger detection of colliding neutron stars through Gravitational Waves (GWs) and Electromagnetic (EM) waves (GW~170817), thanks to the joint efforts of LIGO/Virgo and Space/Ground-based telescopes.
In this paper, we report on the RATIR followup observation strategies and show the results for the trigger G194575. 
This trigger is not of astrophysical interest; however,
is of great interests to the robust design of a followup engine to explore large sky error regions.
We discuss the development of an image-subtraction pipeline for the 6-color, optical/NIR imaging camera RATIR.
Considering a two band ($i$ and $r$) campaign in the Fall of 2015, we find that the requirement of simultaneous detection in both bands
leads to a factor $\sim$10 reduction in false alarm rate, which can be further reduced using additional bands.
We also show that the performance of our proposed algorithm is robust to fluctuating observing conditions,
maintaining a low false alarm rate with a modest decrease in system efficiency that can be overcome utilizing repeat visits.
Expanding our pipeline to search for either optical or NIR detections (3 or more bands), 
considering separately the optical {\it riZ} and NIR {\it YJH} bands, 
should result in a false alarm rate $\approx 1\%$ and an efficiency  $\approx 90\%$.
RATIR's simultaneous optical/NIR observations are expected to yield about one candidate 
transient in the vast 100 $\mathrm{deg^2}$ LIGO error region for prioritized 
followup with larger aperture telescopes.
\end{abstract}

\maketitle

\section{Introduction}
\label{sec:intro}
The first ever direct detection of the GW signal, GW150914, was made by advanced-LIGO \citep{2015CQGra..32k5012A} in September 2015 \citep{2016PhRvL.116m1103A} from a binary BH merger \citep{2016PhRvL.116x1102A}.
This discovery entered us into the GW era; however, complementary identification of  EM counterparts to GW events is required to guide us to the next stage: the GW-EM multi-messenger astronomy era 
\citep{2009arXiv0902.1527B,2010ApJ...725..496N,2012ApJ...746...48M,2016JPhCS.718b2004B}.

A joint EM-GW detection would constrain some fundamental physical properties 
of compact binary coalescence (CBC) events such 
as the distance scale, luminosity, and host galaxy environment. 
However,  identifying a counterpart is remarkably challenging due 
to the LIGO inherently weak localization of GW events ($\sim$ a few hundred $\rm deg^{\rm 2}$).  
Nonetheless, the scientific returns of such discovery justify many efforts taken, even a small step forward. 

CBC events represent powerful engines for the production
of gravitational \citep[see, e.g.,][]{1987thyg.book..330T,1991ApJ...380L..17P,2002ApJ...572..407B,2010CQGra..27q3001A}, 
EM, and neutrino radiation \citep[e.g.,][]{RevModPhys.85.1401}. 
In the CBC model, a neutron star (NS) and compact companion
in an otherwise stable orbit lose energy to gravitational waves 
\citep[e.g.,][]{1987thyg.book..330T,2007PhR...442..166N,2007NJPh....9...17L}. 
Disruption of the NS(s) is thought to produce an accretion disk, which may power relativistic outflows of variable Lorentz factor
\citep{2011ApJ...732L...6R,2016ApJ...824L...6R}. 
Internal shocks in the relativistic jets are expected to produce brief, strong, 
collimated gamma-ray emission and external shocks with the circumstellar 
material are expected to produce the lower-energy emission on longer 
timescales \citep[e.g.,][]{1992MNRAS.258P..41R,1999PhR...314..575P}.
Short-duration Gamma-ray Bursts \citep[sGRBs;][]{1993ApJ...413L.101K} provide our best potential link to gravitational wave sources. 
If these events are due to collapse-object mergers \citep[e.g.,][]{2007PhR...442..166N,2007NJPh....9...17L}, 
copious gravitational waves are expected, 
and these can be detected by Advanced LIGO if the source is sufficiently nearby. 
Indeed, due to beaming \citep{2006ApJ...653..468B,2016ApJ...827..102T}, 
the LIGO rate should be significantly larger 
(factor 10; \citealt{2013PhRvL.111r1101C}) than the observed sGRB rate. 

Finding the potentially rapidly fading afterglow of a GW source requires the engagement of facilities world wide with 
fast response times.
Many facilities have participated in the search for the EM counterparts of LIGO GW events -- in the optical, X-ray,  and radio bands -- and have reported their followup strategies to the community 
\citep[e.g.,][]{2016ApJ...826L...6C,2016MNRAS.460L..40E,2016ApJ...824L..24K,
2016MNRAS.462.4094S,2016ApJ...823L..33S,2016ApJ...828L..16D,2016ApJ...822L...8T}.
We do not expect to see EM emission from BH-BH mergers, so we are mainly
focused on mergers involving at least one NS.

Here, we present the Reionization and Transients InfraRed (RATIR) observatory followup effort.
RATIR is a simultaneous 6-filter imaging camera (\textit{r} band through \textit{H} band), mounted on a 
Harold L. Johnson 1.5-meter telescope
at Observatorio Astron\'omico Nacional on Sierra San Pedro M\'artir, Baja, CA, MX \citep{2012SPIE.8446E..10B,2012SPIE.8444E..5LW}.
The NIR capability of RATIR is highly desirable, with the recent suggestion that some sGRBs 
may be associated with very red ``kilonova" events 
(\citealt{2013ApJ...775...18B,2013Natur.500..547T}, see also \citealt{2015ApJ...811L..22J,2016NatCo...712898J}). 
As an approximately isotropic EM counterpart to the GW signal 
\citep{2009ApJ...691..723B,2011ApJ...736L..21R,2012ApJ...746...48M,2013MNRAS.430.2121P,2017LRR....20....3M}, 
kilonovae provide a unique and direct probe of an important $r$-process site \citep[e.g.,][]{2014MNRAS.439..744R}.
 A kilonova within 100 Mpc would likely
 be quite bright in the NIR and amenable to detection.
 At such distances the source in \citet{2013Natur.500..547T} would have $H < 18.5$ mag (AB).
 RATIR reaches 10$\sigma$ limiting AB magnitudes in 10 minutes of 22.0, 21.4, 20.2, 19.7, 19.6, 18.9 
 in the {\it riZYJH} bands, respectively.

In this paper, we focus our analyses on the trigger G194575 
(\citeauthor[][GCN 18442]{2015GCN..18442...1S}), 
which we were able to promptly observe in Fall 2015
(\citeauthor[][GCN 18455]{2015GCN..18455...1B}).
Despite the fact that this trigger was later found to be unrelated to any astrophysical object later 
(\citeauthor[][GCN 18626]{2015GCN..18626...1L}),
the rather larger error region $\sim$1000 ${\rm deg^2}$ provided us 
a highly challenging exercise for the design of a robust exploratory pipeline.
Similar to other triggers received from the LIGO collaboration team, 
the EM counterpart followup community responded quickly to the trigger 
and was actively engaged until its non-astrophysical origin became apparent.
During that time, ground-based observatories reported two sources of potential interest regarding the trigger, 
LSQ15bjb detected by the La Silla-QUEST 
(\citeauthor[][GCN 18473]{2015GCN..18473...1R})
and iPTF15dld detected by the iPTF 
(\citeauthor[][GCN 18497]{2015GCN..18497...2S}).
RATIR observed the La Silla - QUEST candidate and reported a clear detection of the source in the $i$, $r$, and $z$ bands
(\citeauthor[][GCN 18500]{2015GCN..18500...1G}).

In section 2, we describe the survey strategy, data reduction, and analysis of the designed EM counterparts discovery pipeline.
Field targeting and scheduling along with identification and rejection of bad subtractions are presented in section 3.
In section 4, we discuss our results, the expected false alarm and success rate, 
and address the community benefits from the RATIR pipeline.
All magnitudes in the paper are in the \textit{AB} system.

\section{Survey Strategy, Data Reduction, and Analysis}
\label{sec:mtd}
A search over the entire LIGO detector error region (several hundred square degree) using
a narrow field of view (five to ten arcminute) instrument like RATIR is unfeasible due to
observing time constraints.  It is simply not possible to complete the survey sufficiently rapidly (within a few days)
in two or more epochs to allow for a comprehensive search for variable, new objects.  Instead,
we target only portions of the LIGO error regions most likely to contain sGRBs.

In our strategy (Section \ref{sec:gal}), we search a much smaller portion of the LIGO error region by crossmatching GW galaxy catalog \citep{2011CQGra..28h5016W} sources
and including only very bright luminous galaxies \citep{2016ApJ...820..136G}.
The candidate galaxies selected based on a population-half-light criteria using the absolute $B$-band magnitudes, 
and are scheduled for visits twice per field. 
Return visits for image subtraction purposes are conducted on a subsequent nights.

With two or more frames captured for each target galaxy field, we performed digital image subtraction with the
High Order Transform of PSF ANd Template Subtraction
\citep[HOTPANTS;][]{2015ascl.soft04004B}. 
HOTPANTS is an implementation of the \citet{1998ApJ...503..325A} algorithm,
based on a spatially-variable kernel method that matches the PSFs of two astronomical images.
Prior to running, the images are bias, dark, and sky-subtracted, flat-fielded, and astrometrically co-aligned
using SWARP \citep{2002ASPC..281..228B}.
We use the Source Extractor \citep{1996A&AS..117..393B} software to identify sources for alignment and to estimate
the image FWHM values.  The quadrature difference between FWHM values is used to define the starting Gaussian sigma values
for the HOTPANTS convolution.  Custom point-spread-function (PSF) fitting software is used to estimate the image PSF and
to obtain photometry for the difference frames.  In our final photometric detections, we require $\ge 10\sigma$ detections.

While our image subtractions are typically very clean (e.g., Figure \ref{fig:img_sub_gal}), residual flux
can often be detected near bright sources or new image boundaries.  These false sources are flagged and ignored (see Figure \ref{fig:fdrmstp})
by identifying a bright cataloged source within 10 arcsec, by comparing the (typically small)
FWHM of the source relative to the median FWHM of the image, or by discarding sources near the image boundaries.
Bad subtractions can also be obtained, typically yielding a large number ($>10$) of detections.  We have developed
an automated filtering approach to minimize these cases using image quality metrics present prior to subtraction (Section \ref{subsec:ign}).

We distinguish among three types of false alarms: 
non-astrophysical fake signals (largely from bad subtractions); 
known asteroids and other known solar system bodies; 
and astrophysical transients that do not correspond to the GW events. 
In this work, we focus on filtering out the first type.

\begin{figure}[!ht]
\centering
\subfigure[{\scriptsize }]{
\includegraphics[width=\ww\columnwidth, keepaspectratio]{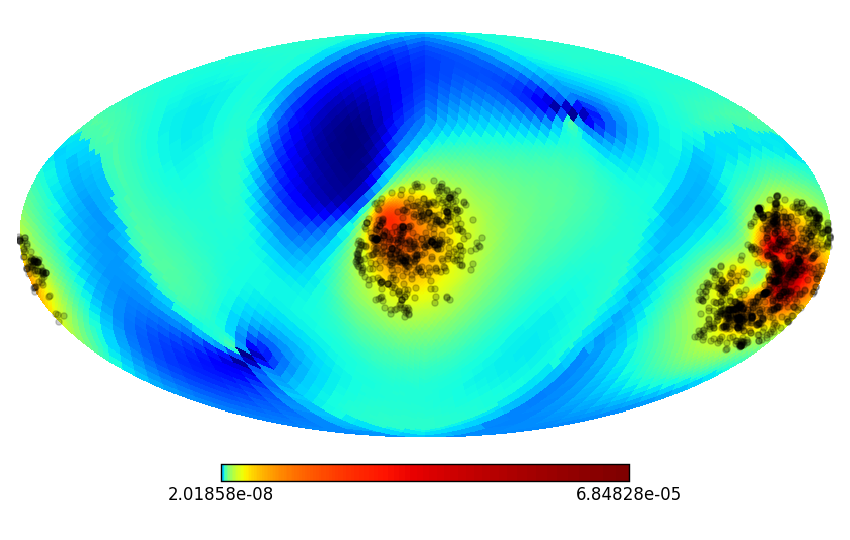} 
	\label{fig:subfig1.1} 
}
\\
\subfigure[{\scriptsize }]{
\includegraphics[width=\ww\columnwidth, keepaspectratio]{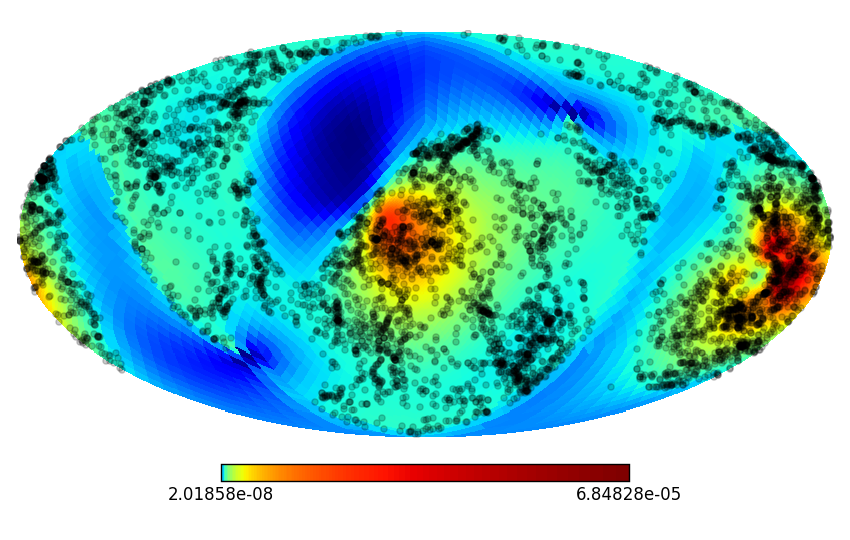}      
\label{fig:subfig1.2}  
}
\\
\subfigure[{\scriptsize }]{
\includegraphics[width=\ww\columnwidth, keepaspectratio]{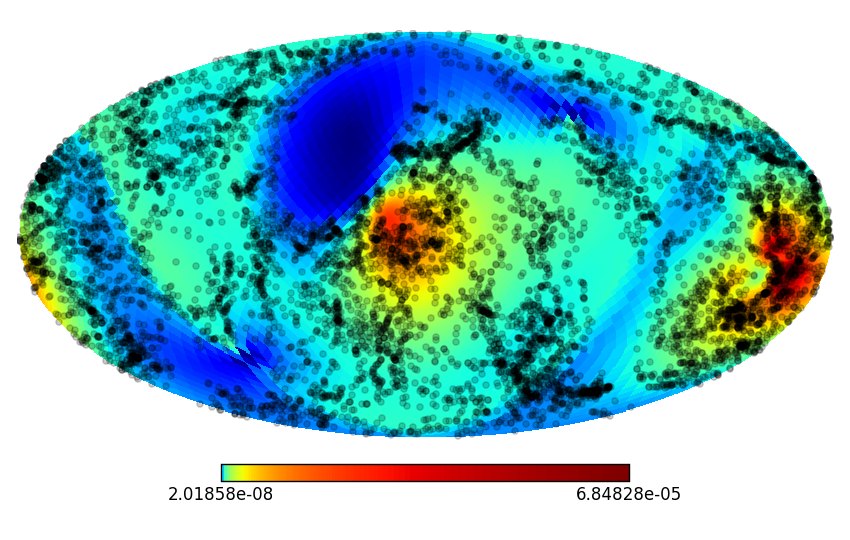}     
\label{fig:subfig1.3}  
}
\vspace{-10pt}
\caption{
The BAYESTAR GW probability skymap in \textit{log} for trigger G194575 
over-plotted with the list of galaxies in the GWGC with black dots
for 1$\sigma$ (fig. 1a), 2$\sigma$ (fig. 1b), and 3$\sigma$ (fig. 1c). 
}
\label{fig:sky_map}
\end{figure}

\subsection{Galaxy Strategy}
\label{sec:gal}

A typical LIGO sky map error region is much larger than the field of view of optical or X-ray telescopes. 
Due to the time constraints of rapid followup, covering a few hundred or even a few tens of the sky square degrees is not a practical approach by a small FoV telescope. 
Therefore, finding an optimal strategy which determines the ideal domain of investigations should be at the core of any pipeline designs of LIGO GW followup sources \citep[e.g.,][]{2014ApJ...784....8H,2015ApJ...801L...1B}.  
A catalog of galaxies which has already satisfied some critical criteria relevant to our search is required. 
These criteria are   
adequate sky coverage,  sufficient depth, and galaxy brightness (high blue luminosity).
The latter condition is important because blue luminosity is a tracer of recent star formation and sources produced by stars ought to track the light.

The Gravitational Wave Galaxy Catalog \citep[GWGC;][]{2011CQGra..28h5016W} is an attempt to offer such a galaxy catalog and has been used by many followup groups e.g. DLT40, Swift-XRT, UL50, Kanata, OAO-WFC, and RATIR. 
We used GWGC as our main catalog during \textit{O1} while also following a similar galaxy strategy as \citet{2016ApJ...820..136G} which considers only brighter galaxies that produce $\sim$50\% of the light. This constrains the absolute blue magnitude of galaxies to less than -20.025 mag; eliminating $\sim$80\% of galaxies in the GWGC.  

The modified GWGC is complete $\sim$100\% out to about 60 Mpc which is consistent with the LIGO estimate of sensitivity coverage during \textit{O1} run.
The estimate of completeness is defined based on $B$-band magnitude which is expected to follow sGRB rate \citep{2013ApJ...769...56F}.
At distances $\sim$100 Mpc, the GWGC completeness reduces to 
about 85\% for the selected bright galaxies (see the figure 3 of \citealt{2016ApJ...820..136G}). 
Caution will be necessary during the LIGO \textit{O2} and later runs since we expect to detect GW events due to the binary NS at distances exceeding 100 Mpc \citep{2016LRR....19....1A}, beyond which the incompleteness of the GWGC increases.
New catalogs more suitable for the next LIGO runs are under construction (e.g. CLU, \citealt{2016ApJ...820..136G}, and GLADE\footnote{\url{http://aquarius.elte.hu/glade/}}, \citealt{2016yCat.7275....0D}). 

\section{Field Targeting and Scheduling}
\label{sec:anlaz}

Upon a trigger, our pipeline automatically receives the probability skymap error region from the LIGO collaboration. The sky localization is provided at low-latency by the ``BAYESTAR'' and ``CWB'' pipelines, and later with ``LALInference''.
The skymap error region is projected onto the modified GWGC, as described in the previous section, 
and a list of candidate galaxies is made. 
Figure \ref{fig:sky_map} shows the BAYESTAR GW probability skymap  
for trigger G194575 (\citeauthor[][GCN 18442]{2015GCN..18442...1S}).
Regions with the darker red color represent higher probability GW source localization and regions with the darker blue color are associated with the lowest probability GW source locations. 
The color bar shows the corresponding probability values.
The GWGC contains 53,312 galaxies and our bright galaxy criterion ($x_{1/2} > 0.626$,  see \citet{2016ApJ...820..136G} for details) typically passes only $\sim$20\% in a given sky area.
Projecting the G19475 skymap within 1$\sigma$ error region onto our galaxy catalog results in 1539 candidate galaxies which are shown with black circles in Figure 1a. 
This number increases to 6057 and 8217 for 2$\sigma$ (fig. 1b) and 3$\sigma$ (fig. 1c), respectively.

Given that not all of these galaxies will be visible at OAN-SPM, 
we expect to followup about half of the galaxies in the list. This is still a large number even for the 1$\sigma$ error region. 
The RATIR scheduler selects and queues targets automatically 
to be imaged at the first available time.
Nominally, we also rank the list of candidate galaxies based on the $B$-band luminosity value. 
It is also possible to prioritize based on distance estimates provided in the LIGO/Virgo GCN notices.
However, an estimate of the LIGO GW source distance was not available during
the \textit{O1} run.
 
Given the available observing time, we observed 26 nearby galaxies ($D<10$ Mpc) 
within the GWGC catalog and contained with the 1$\sigma$ LIGO/Virgo error 
region for trigger G194575 (\citeauthor[][GCN 18442]{2015GCN..18442...1S}).
Between RA 23.1 hours and RA
1.5 hours (J2000) with RATIR on the night of 2015/10/23,
we obtained a total exposure of $\sim$8 minutes on each of the 26 fields (see Table 1), reaching typical
depths of $r$ and $i = 21$ mag (AB, 10$\sigma$).  These magnitudes
are not corrected for Galactic extinction.  Each
field, centered upon one GWGC galaxy, has a size of approximately $5 \times 5$ 
$\rm arcmin^2$.
We observed these fields on two more consecutive nights (10/24 and 10/25; Table 1).
We had a typical image quality of 1.8 arcsec each night.  
To reach comparable limiting magnitudes on the third and final night, a 50\% 
increase in exposure time was required to overcome highly non-photometric observing conditions.  
On the first two nights, the $i$-band zero points were stable (to within 10\%) 
while the zero point varied by nearly a factor of unity on the third night.
The NIR RATIR channels ({\it ZYJH}) were not available at this time.

\begin{figure*}[!ht]
\vspace{-5pt}
\centering
\textbf{\qquad\;\;\;Science \qquad\qquad\qquad\qquad\;\; Reference \qquad\qquad\qquad\qquad\qquad\;\; Subtracted}\par
\subfigure{
\rotatebox{90}{$\qquad\qquad\;\;\; \textbf{\textit{i}-band}$}
\hspace{5pt}
\includegraphics[width=\vvw\columnwidth, keepaspectratio]{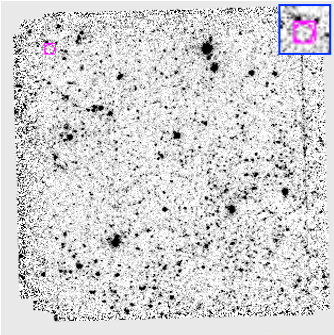} 
\label{fig:subfig1} 
}
\subfigure{
\includegraphics[width=\vv\columnwidth, keepaspectratio]{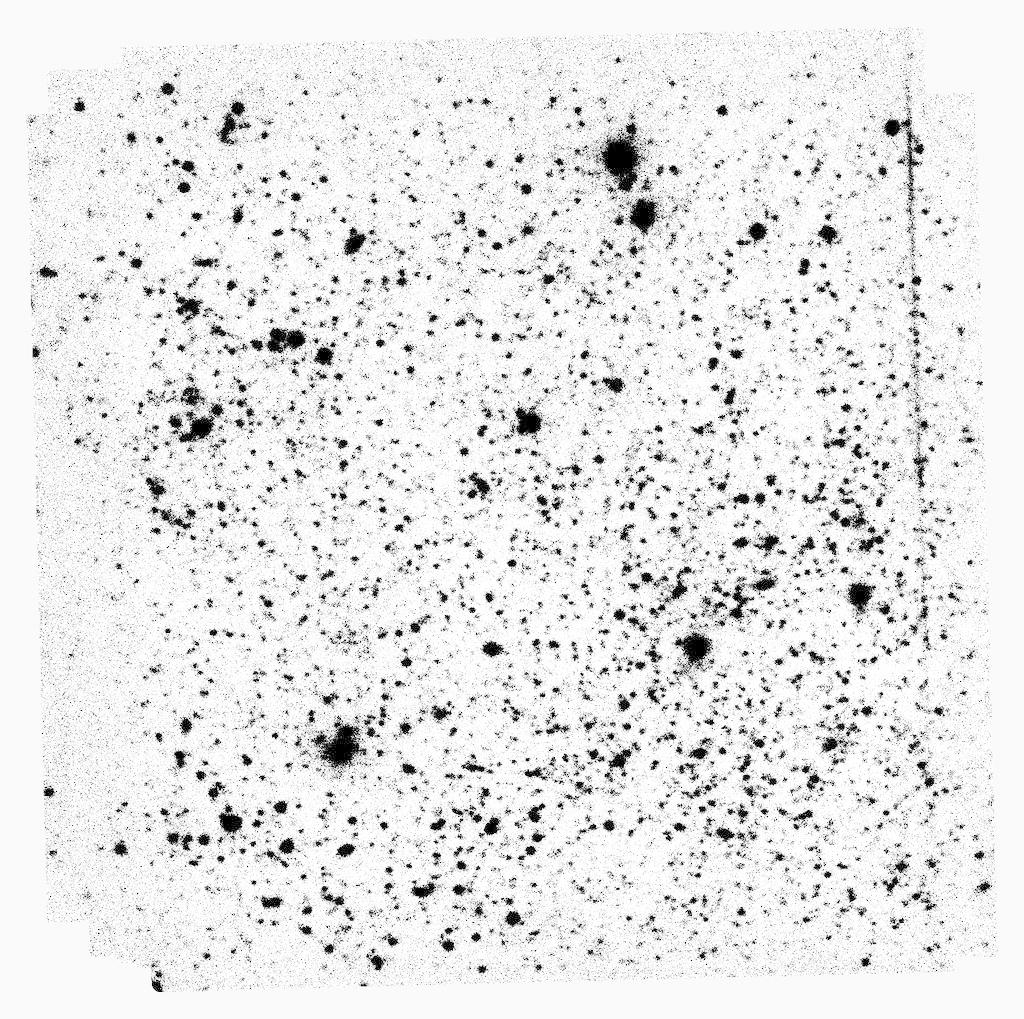}    
\label{fig:subfig2}  
}
\subfigure{
\includegraphics[width=\vvw\columnwidth, keepaspectratio]{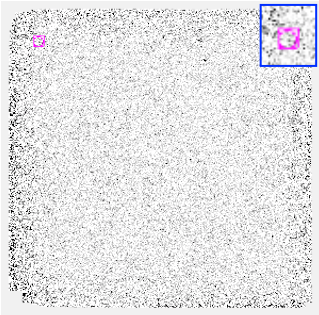}      
\label{fig:subfig3}  
}
\\
\vspace{-20pt}
\subfigure{
\rotatebox{90}{$\qquad\qquad\;\;\; \textbf{\textit{r}-band} \qquad\qquad\quad \underline{\#14}$}
\hspace{5pt}
\includegraphics[width=\vv\columnwidth, keepaspectratio]{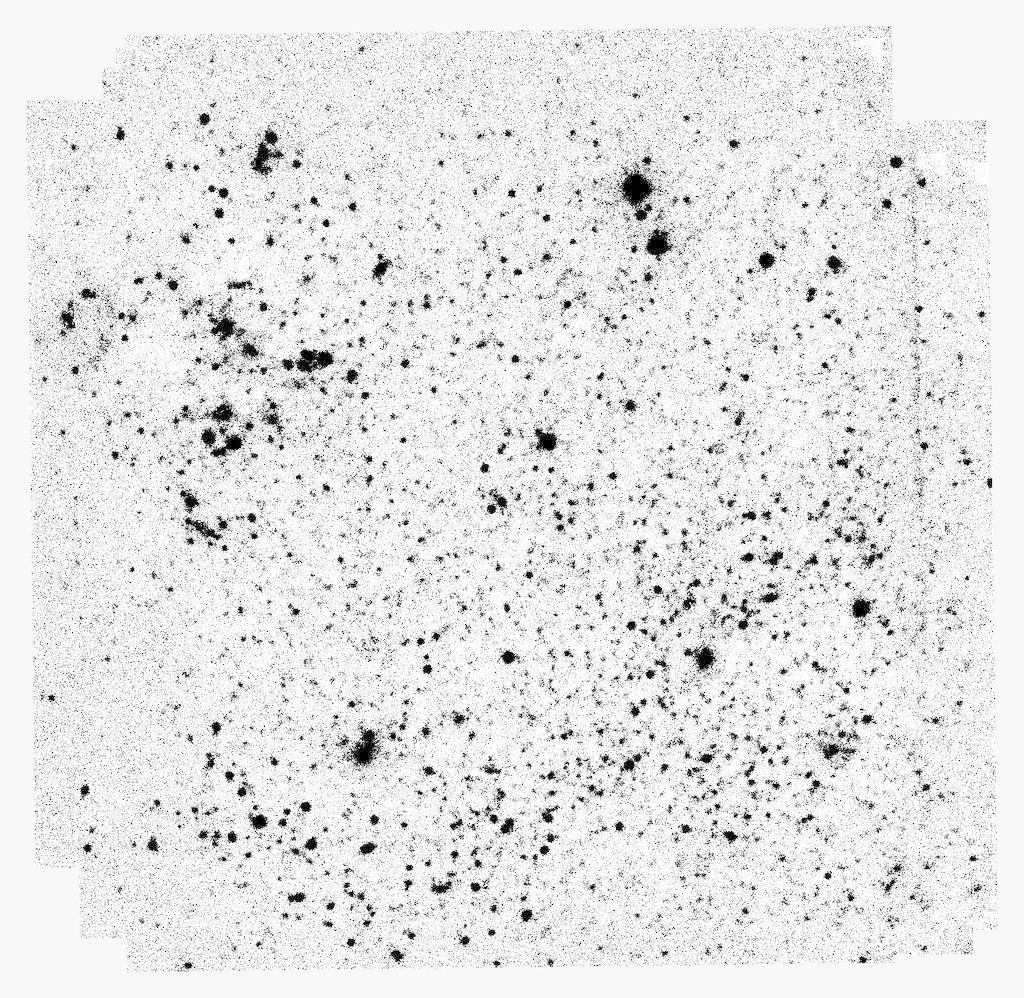}    
\label{fig:subfig10}
}
\subfigure{
\includegraphics[width=\vv\columnwidth, keepaspectratio]{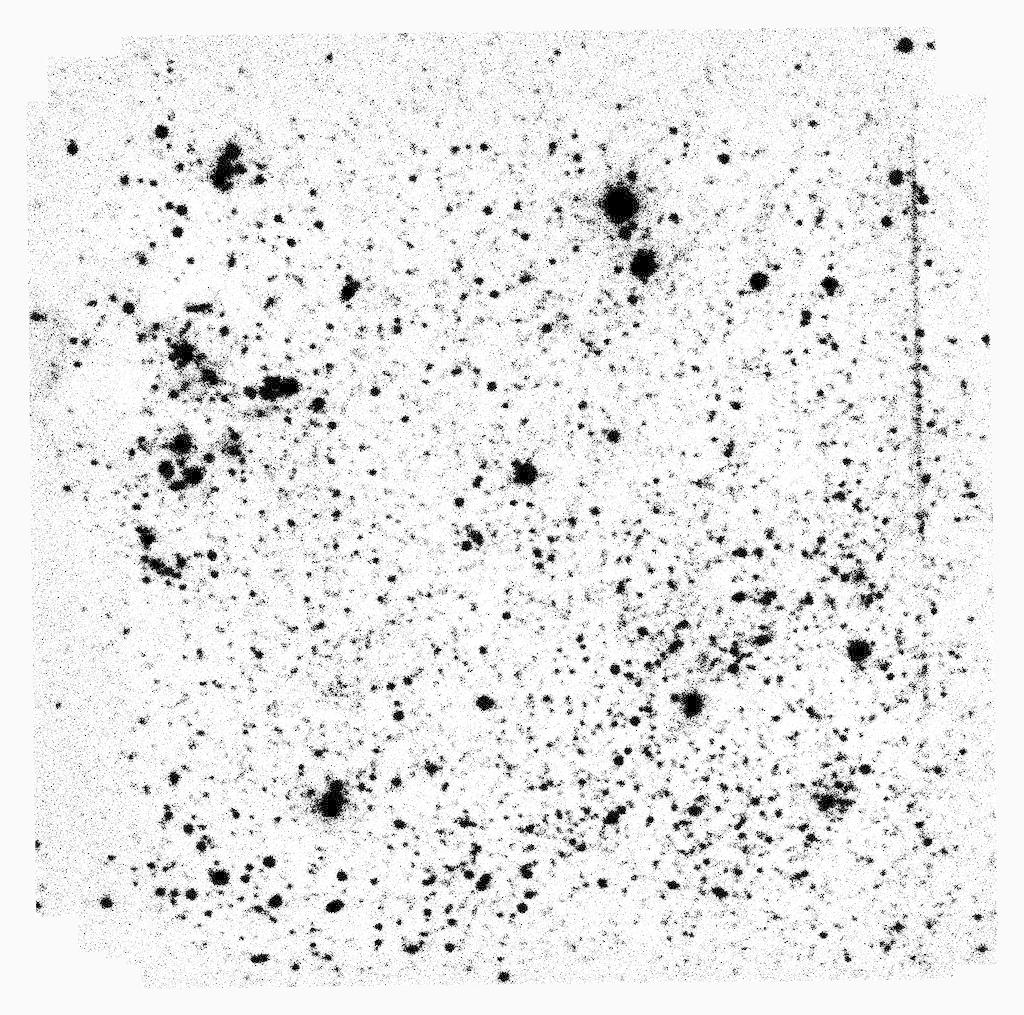}    
\label{fig:subfig11}
}
\subfigure{
\includegraphics[width=\vv\columnwidth, keepaspectratio]{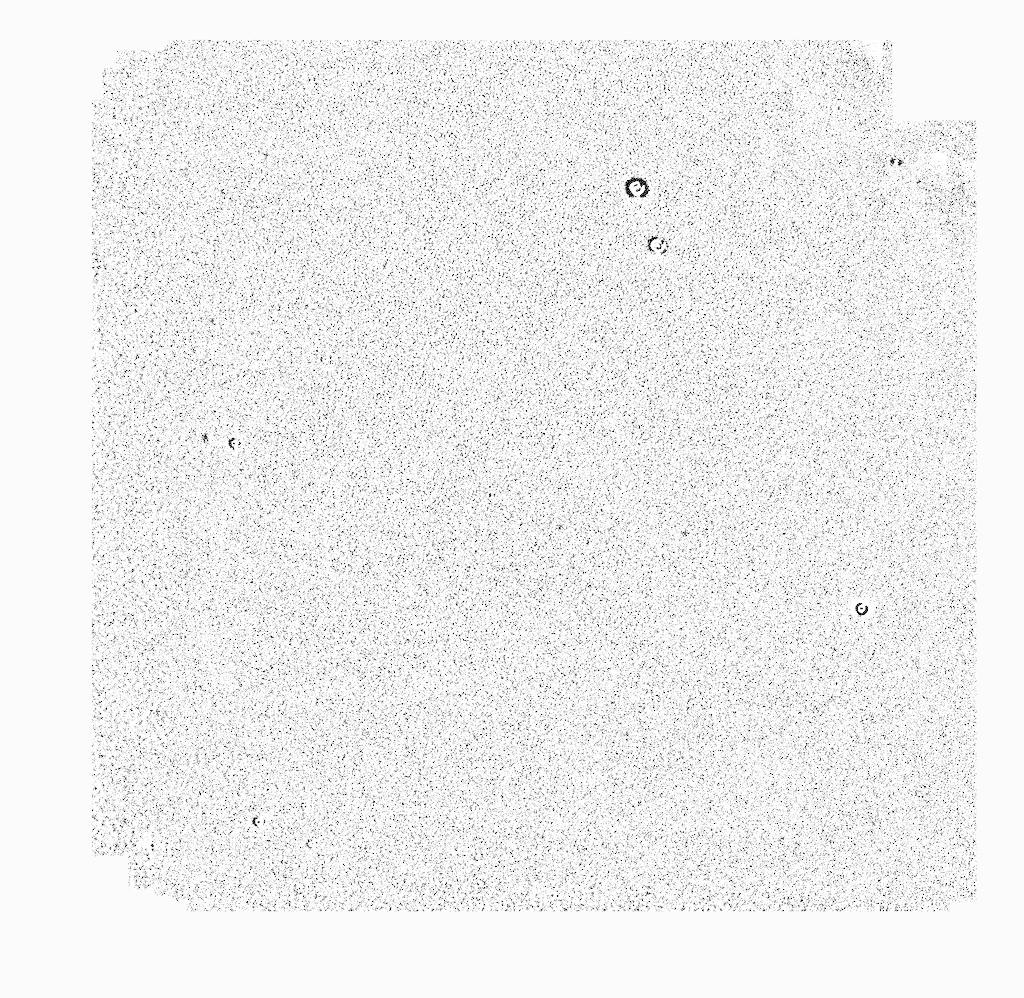}    
\label{fig:subfig12}
}
\\
\subfigure{
\rotatebox{90}{$\qquad\qquad\;\;\; \textbf{\textit{i}-band}$}
\hspace{5pt}
\includegraphics[width=\vvv\columnwidth, keepaspectratio]{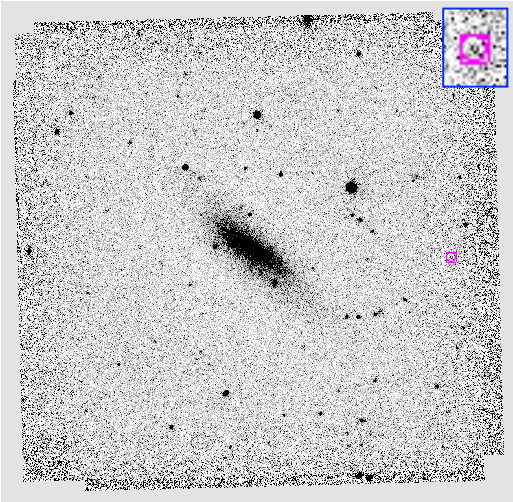}    
\label{fig:subfig4}
}
\subfigure{
\includegraphics[width=\vv\columnwidth, keepaspectratio]{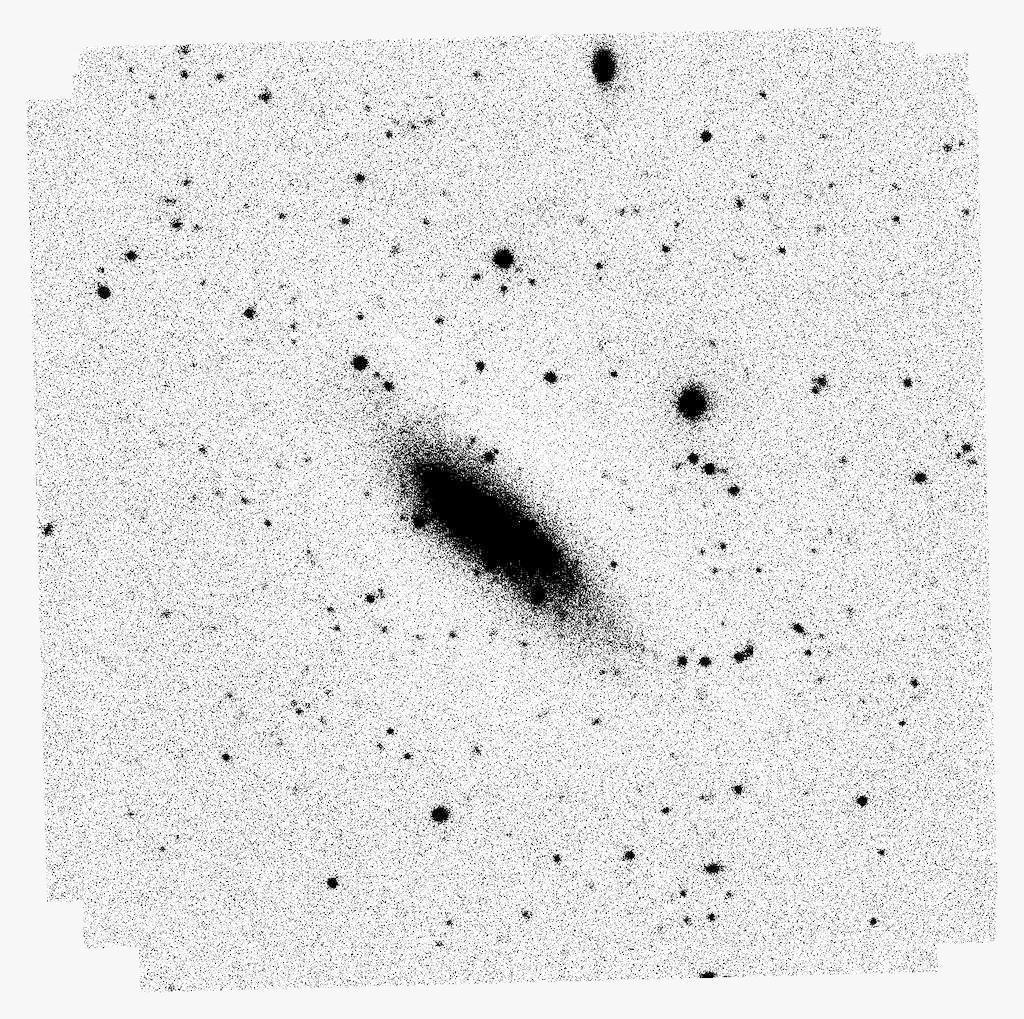}    
\label{fig:subfig5}
}
\subfigure{
\includegraphics[width=\vvv\columnwidth, keepaspectratio]{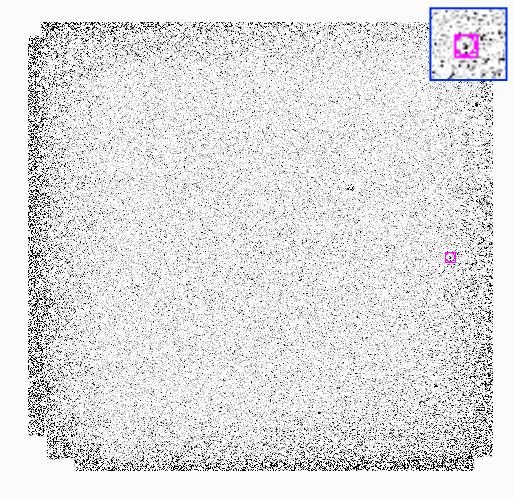}    
 \label{fig:subfig6}
}
\\
\vspace{-20pt}
\subfigure{
\rotatebox{90}{$\qquad\qquad\;\;\; \textbf{\textit{r}-band} \qquad\qquad\quad \underline{\#16}$}
\hspace{5pt}
\includegraphics[width=\vv\columnwidth, keepaspectratio]{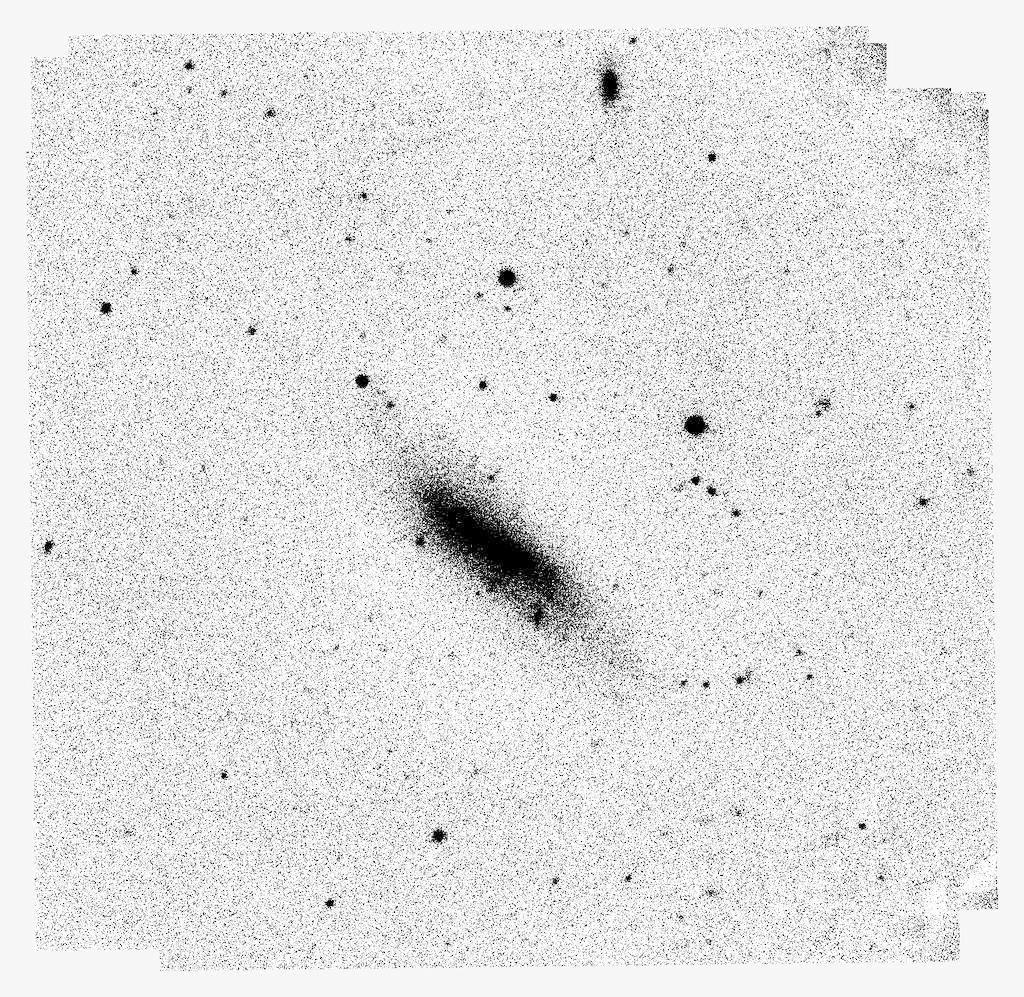}    
\label{fig:subfig13}
}
\subfigure{
\includegraphics[width=\vv\columnwidth, keepaspectratio]{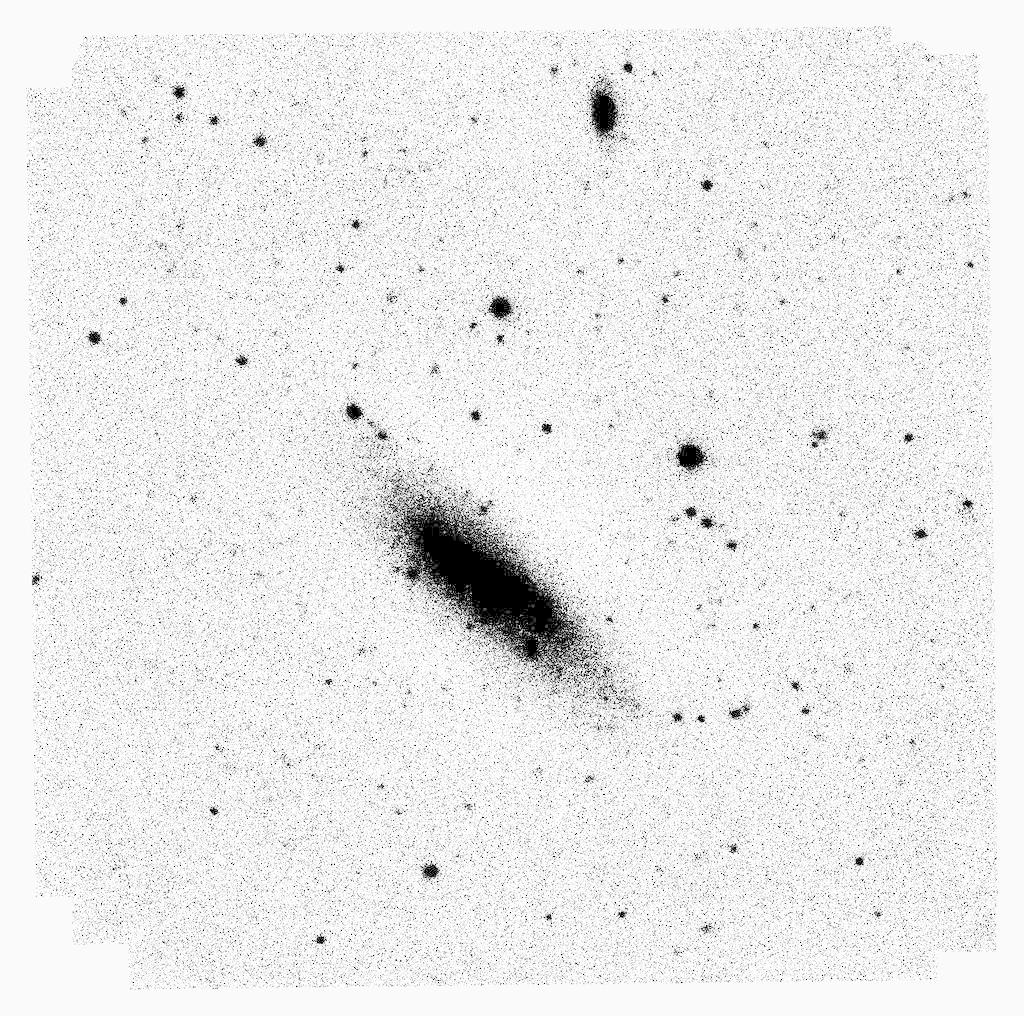}    
\label{fig:subfig14}
}
\subfigure{
\includegraphics[width=\vv\columnwidth, keepaspectratio]{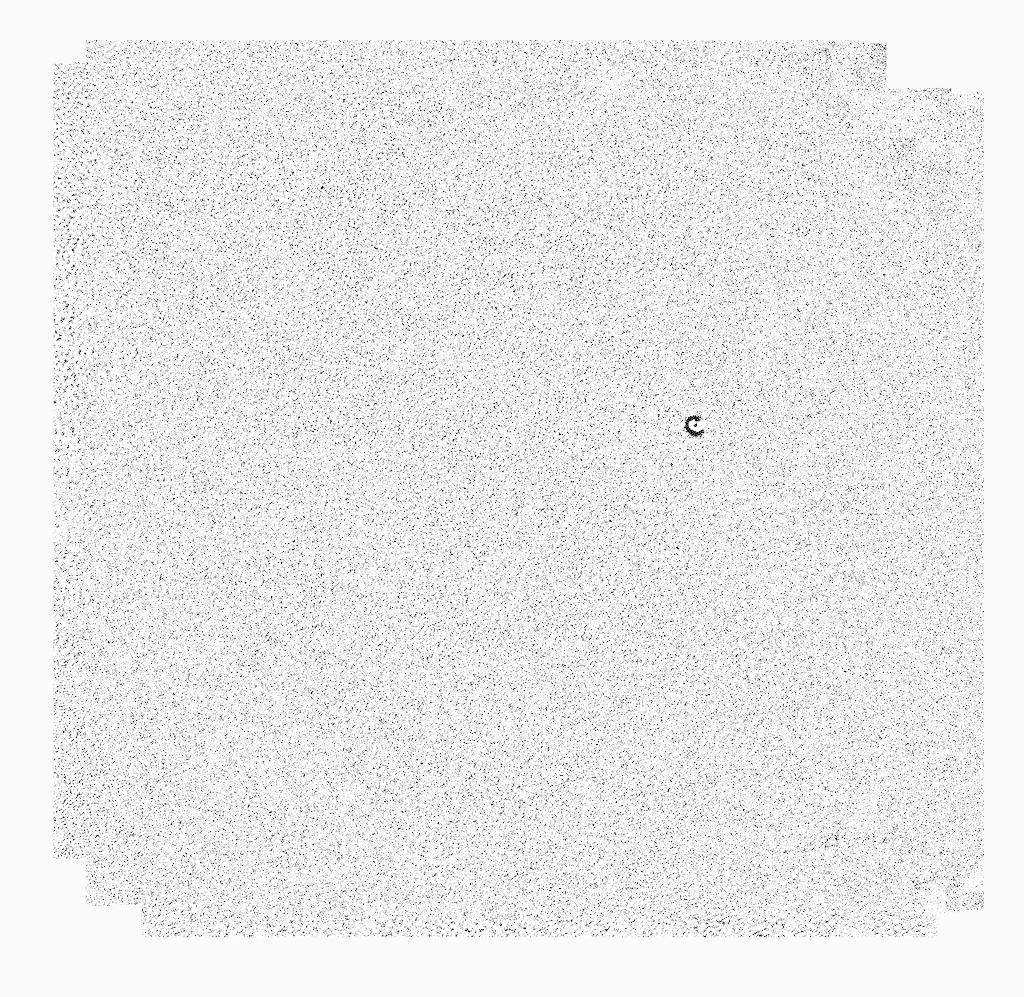}    
\label{fig:subfig15}
}
\\
\subfigure{
\rotatebox{90}{$\qquad\qquad\;\;\; \textbf{\textit{i}-band} $}
\hspace{5pt}
\includegraphics[width=\vvv\columnwidth, keepaspectratio]{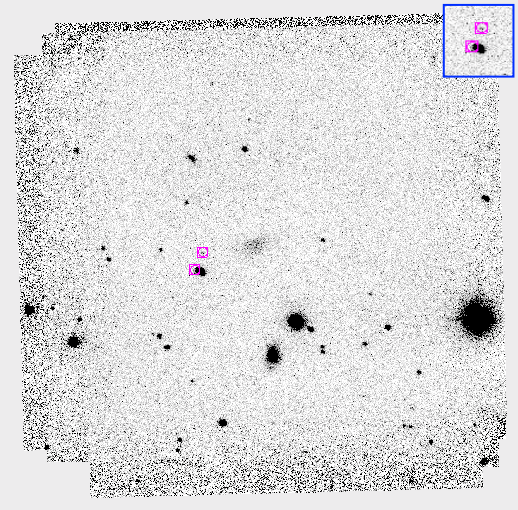}    
\label{fig:subfig7}
}
\subfigure{
\includegraphics[width=\vv\columnwidth, keepaspectratio]{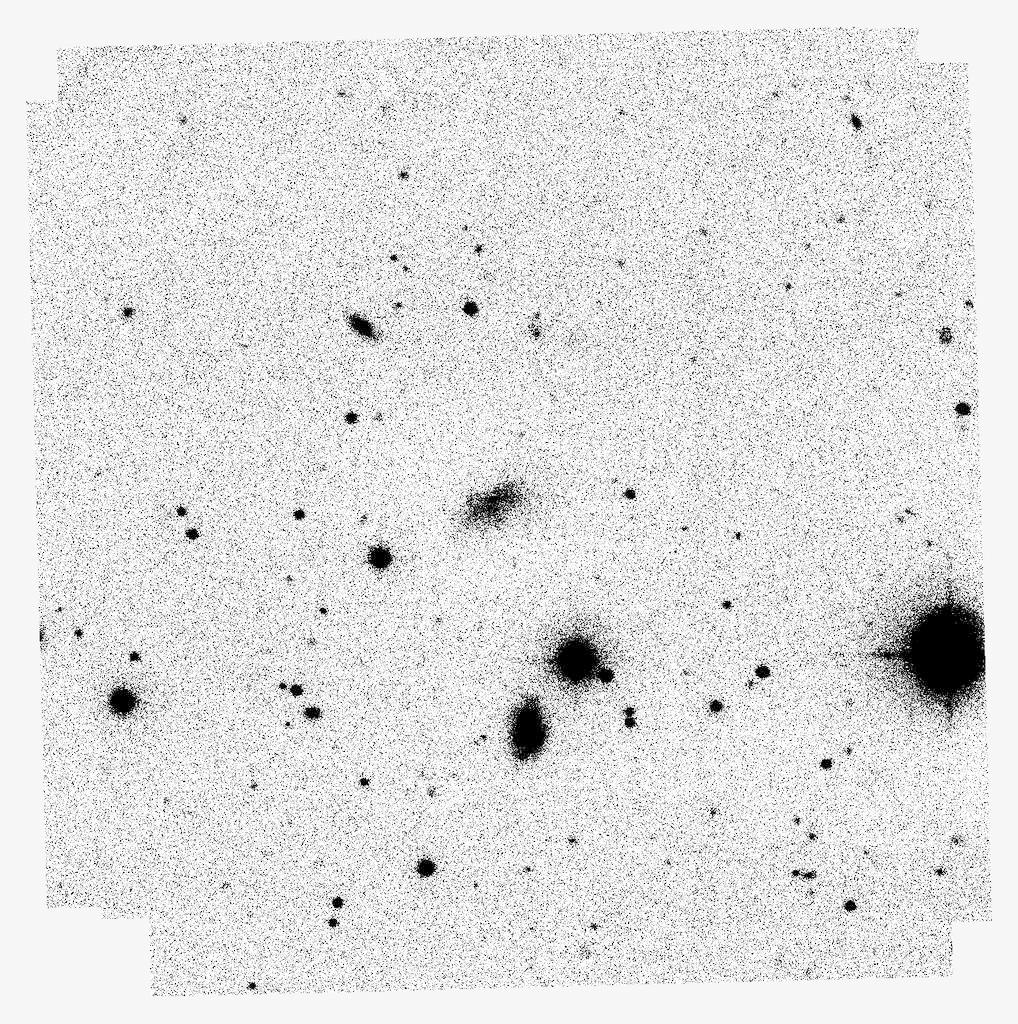}    
\label{fig:subfig8}
}
\subfigure{
\includegraphics[width=\vvv\columnwidth, keepaspectratio]{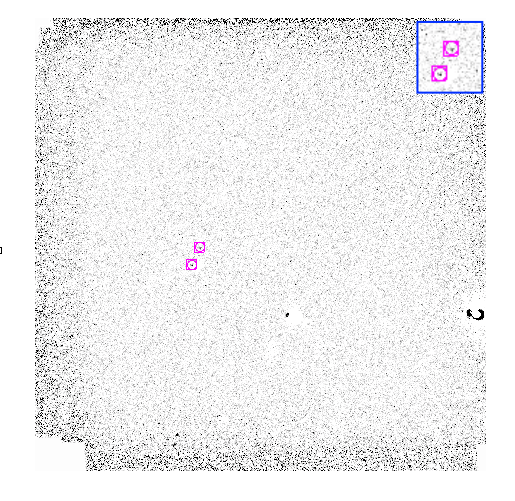}    
\label{fig:subfig9}
}
\\
\vspace{-20pt}
\subfigure{
\rotatebox{90}{$\qquad\qquad\;\;\; \textbf{\textit{r}-band} \qquad\qquad\quad \underline{\#23}$}
\hspace{5pt}
\includegraphics[width=\vvv\columnwidth, keepaspectratio]{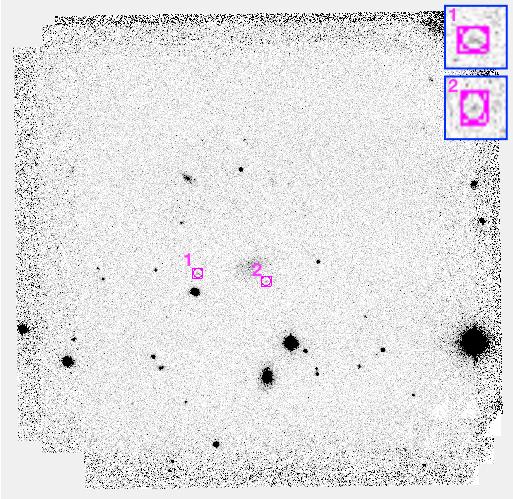}    
\label{fig:subfig16}
}
\subfigure{
\includegraphics[width=\vv\columnwidth, keepaspectratio]{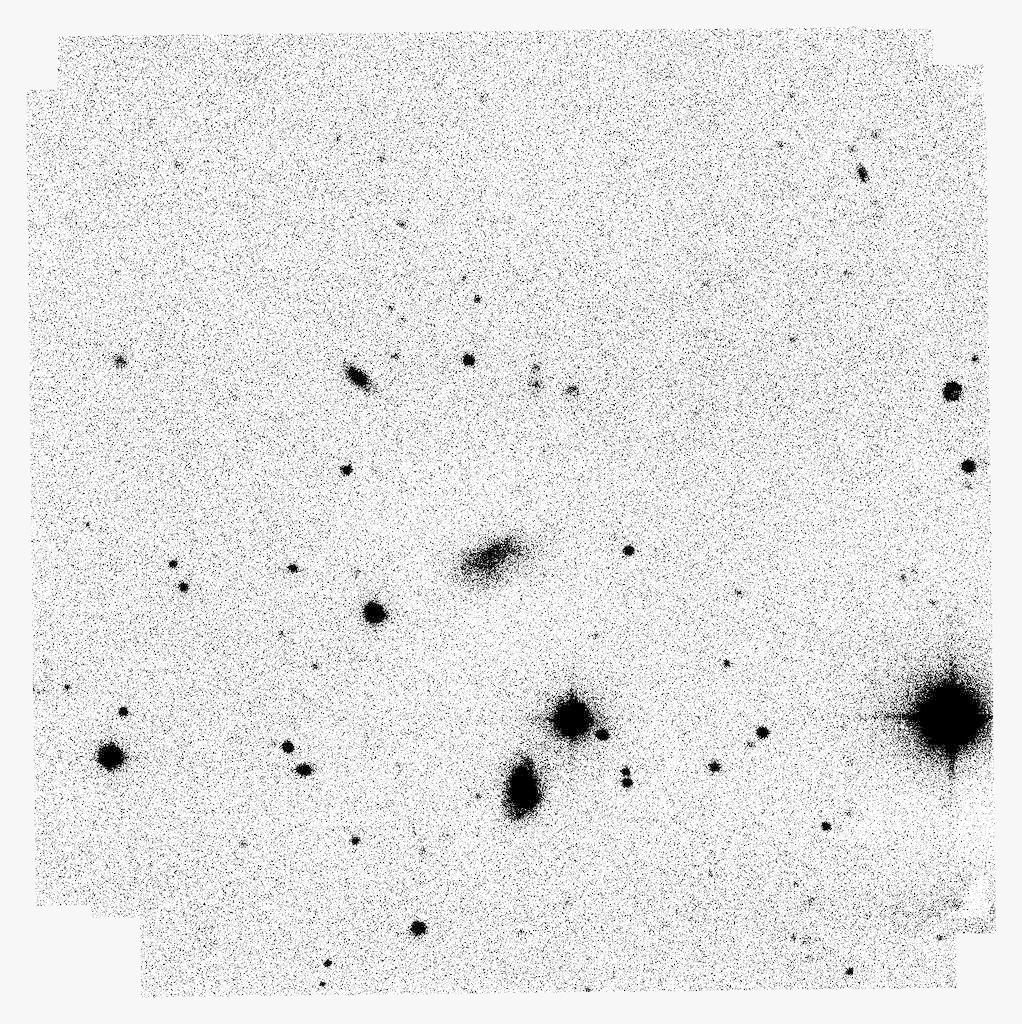}    
\label{fig:subfig17}
}
\subfigure{
\includegraphics[width=\vvv\columnwidth, keepaspectratio]{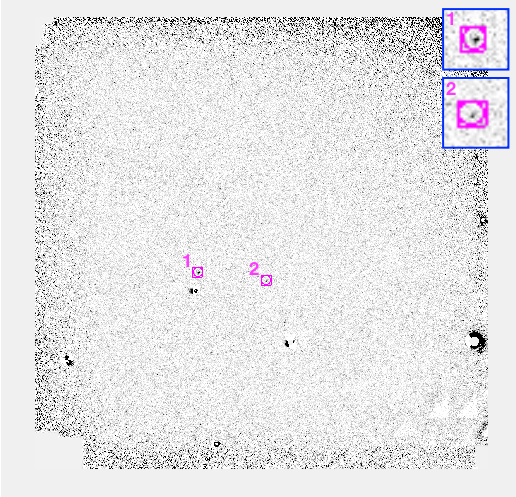}    
\label{fig:subfig18}
}
\vspace{-12pt}
\caption{
A gallery of image subtraction frames for targets
\#14 ($1^{\rm st}$ \& $2^{\rm nd}$ rows), \#16 ($3^{\rm rd}$ \& $4^{\rm th}$ rows), 
and \#23 ($5^{\rm th}$ \& $6^{\rm th}$ rows) in the $i$ and $r$ bands.
The images on the left, middle, and right show science (10/24), reference (10/23), and
the subtracted frames, respectively. 
Detected sources marked with red squares in the science and subtracted frames.
The sizes of all the postage-stamps are: $5 \times 5$ $\rm arcmin^2$.
}
\label{fig:img_sub_gal}
\end{figure*}

\begin{figure*}[!ht]
\centering
\includegraphics[width=1.\columnwidth]{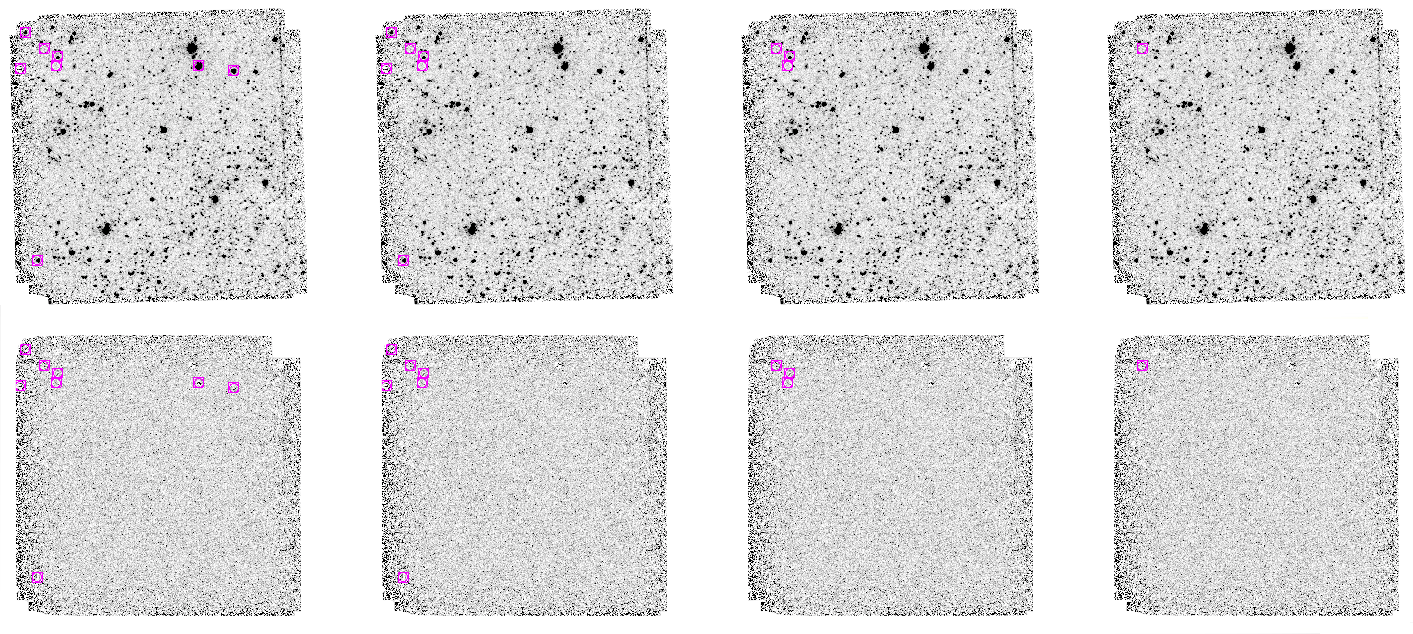}
\vspace{-10pt}
\caption{\small 
Various steps of removing false detected sources marked with red squares of target \#14 in 
the source frame $i$-band (top row) and the subtracted frame (bottom row).
All the raw detected sources are shown in the first column (no filter).
The initial step of filtering rejects multiple detections around same stars (second column).
The subsequent step of filtering removes false detections on the edge of a frame (`distance to edge' $< 62$ pixel).
The last step of filtering rejects extra clustered detections around a source (`cluster radius' $< 14$ arcsec).
The sizes of all the postage-stamps are: $5 \times 5$ $\rm arcmin^2$.
}
\label{fig:fdrmstp}
\end{figure*}

Figure \ref{fig:img_sub_gal} shows a gallery of image subtraction frames for targets
\#14, \#16, and \#23 in the $i$ and $r$ bands.
The images on the left, middle, and right columns show science, reference, and
the subtracted frames, respectively. 
We note that the image subtraction is typically extremely clean.
Each of the selected targets shown in the Figure \ref{fig:img_sub_gal} represents
a different type of target field. Target \#14 is a very crowded field with many stars in the foreground
and thus not very deep. Target \#16 is a deep image with the targeted galaxy in the center
and adequate sources for the alignment task, ideal for our purpose. 
Target \#23 is a semi-crowded field with manageable number of sources for image-subtraction.  
Detected sources are marked with red squares in the science and subtracted frames.
Fields \#14 and \#16 contain only one false detection in the $i$ band. 
In field \#23, each of the $i$ and $r$ bands comprises two false detections; however, 
only one of them appears in the same position on the two frames. 

Observations obtained on the first night are compared with SDSS images, if available, 
as reference frames to detect any potential targets. 
The RATIR pipeline starts observing the LIGO/Virgo error region immediately 
after receiving a trigger form the LIGO team. 
Any meaningful changes in the magnitude of a true candidate source 
should evolve during the consecutive nights in which we are able to detect.
Given the requirement of $10\sigma$ threshold detection in our designed pipeline, 
any changes of magnitude $\gtrsim 0.1$ mag should be detected. 

\subsection{Identifying and Ignoring {\it Bad} Subtractions}
\label{subsec:ign}

We now implement our modified image-subtraction routine (Section \ref{sec:mtd})
to detect any possible transients (as an example, see Figure \ref{fig:fdrmstp}).
The top row in Figure \ref{fig:fdrmstp} shows target \#13 (science frame from 10/24) and the bottom row shows the subtracted frame (the 10/23 frame was used as the reference frame). 
Each column shows various steps of removing false detections.
No filter is implemented for the first column. The initial step of filtering rejects multiple detections around same stars (second column).
The subsequent step of filtering removes false detections on the edge of a frame (`distance to edge' $< 62$ pixel). 
The last step of filtering rejects extra clustered detections around a source (`cluster radius' $< 14$ arcsec).
The implemented actions, the presented hierarchical steps, 
and the numbers described here are the outcome of a comprehensive 
statistical analysis with the objective of reducing image subtraction residuals.

In principle, having
observed 26 galaxy targets in 2 optical bands for 3 consecutive nights, we can carry-out
$26 \times 3 \times (2) \times 2 = 312$ different image subtractions.
The extra factor of 2 takes into account the two possible way of subtracting two images.
In practice, we would like to only conduct one subtraction per field in a way that yields the
highest quality subtraction.  We now use all possible subtractions to explore how to find the best possible subtraction.

\begin{figure}[!ht]
\centering
\includegraphics[width=2.8in]{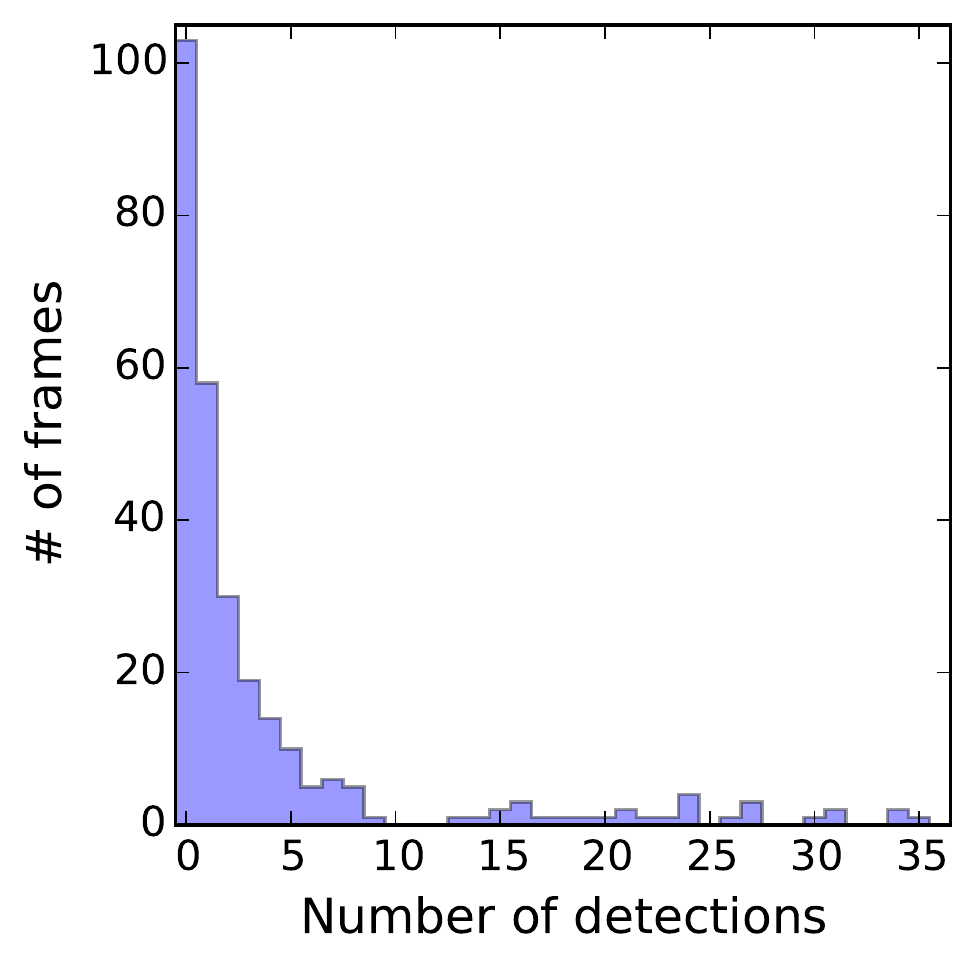}
\caption{\small 
Histogram of number of detections in all frames.
}
\label{fig:hist}
\end{figure}

Figure \ref{fig:hist} shows a histogram of the number of detections in all frames.
The number of false detections is zero for about half of the frames. For 90\% of the rest is less than 10 detections, and we take $<10$ detections to define a good subtraction.
A well-posed subtraction will always have a deeper reference frame with better seeing as compared to the science frame.  However, because our (time-limited) observation strategy typically leads to similar depths for both {\it science} and {\it reference} frames, it is often not possible (e.g. due to changing sky transmission) to know a-priori how to do the subtraction.  Nevertheless, image statistics determined pre-subtraction can help us to address this problem. 

We exploit the following image-quality statistics and seek to understand how these can be used to avoid bad subtractions:
(1) the ratio of the PSF sigma in the science frame over the PSF sigma in the template, 
(2) the ratio of number of stars in science and template frames, and 
(3) the median of the RMS-fraction between science and template frames.
Hereafter, we refer to these parameters as {\it Par$_1$}, {\it Par$_2$}, and {\it Par$_3$}, respectively.
We now seek to define a sequential filtering on these parameters -- to be coded into an image subtraction wrapper -- that
yields the most compact $K$\%-Confidence Interval (C.I.; Eq. \ref{eq:CI}), e.g. 90\%, dumping outliers).

\be
\mathrm{C.I.} = P(a_k \leq \mathrm{Par_k} \leq  b_k) 
\label{eq:CI}
\ee
Here subscript $k=[1,2,3]$ and the $a$ and $b$ values specify the parameter 
lower and upper bounds that contain C.I. percent of that parameter space.

\begin{figure*}[!ht]
\centering
\includegraphics[width=1\textwidth]{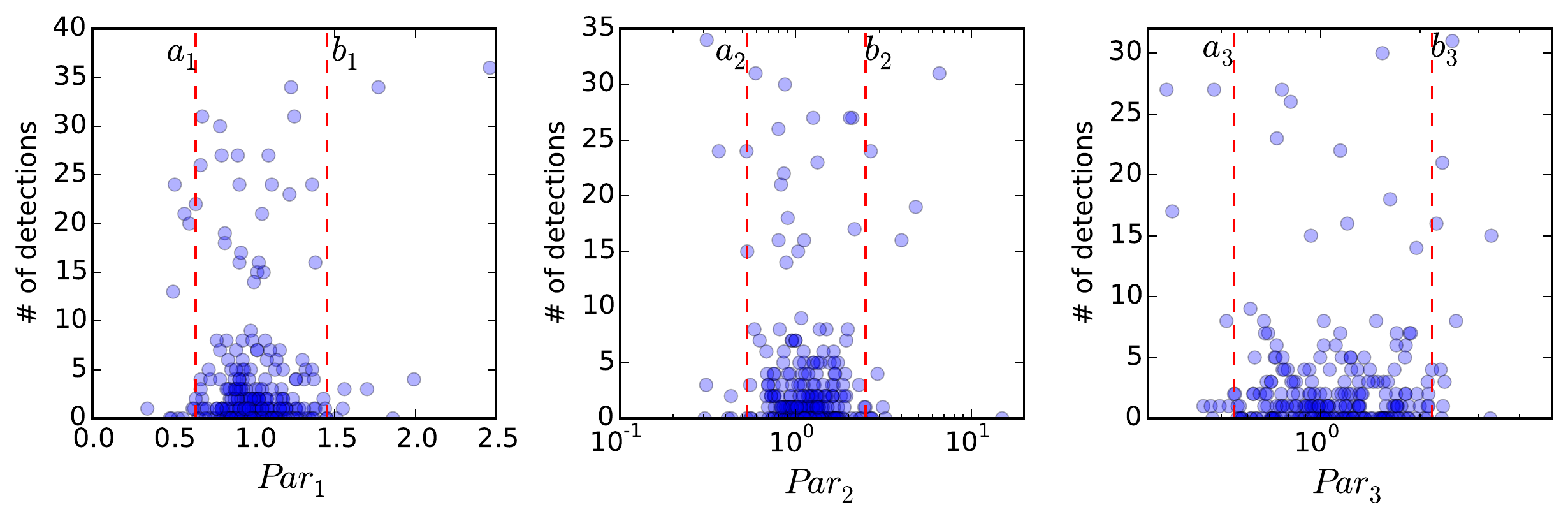} 
\vspace{-12pt}
\caption{\small 
Three stages of filtering implementing consecutively on the {\it Par$_1$} 
($\sigma_{\rm sci}/\sigma_{\rm tmp}$), 
{\it Par$_2$} ($N^{\star}_{\rm sci}/N^{\star}_{\rm tmp}$), 
and {\it Par$_3$} ($R\!M\!S_{\rm sci}/R\!M\!S_{\rm tmp}$) 
parameter spaces, in order. Cuts in the parameter space keep data within the most 
compact 90\%-confidence interval and are shown with dashed lines.
}
\label{fig:3dflt}
\end{figure*}

Figure \ref{fig:3dflt} shows scatter plots of the {\it Par$_1$}, {\it Par$_2$}, and {\it Par$_3$} values of all subtracted images versus their number of detected sources, separately. 
It also presents the three stages of filtering implementing consecutively on their parameter spaces, in the same order. 
The {\it Par}$_1$ data within $[a_1, b_1]$ range feeds to the second filter which implements on the {\it Par}$_2$ space, and so forth, the {\it Par}$_2$ data within $[a_2, b_2]$ range feeds to the third filter which implements on the {\it Par}$_3$ space. To reach the 90\%-C.I. at the end of filtering task, the image subtraction wrapper figures out how to set the same value of $K$ between the filtering stages.
Cuts in the parameter space are shown with dashed lines in the Figure \ref{fig:3dflt}.

\begin{figure}[!ht]
\centering
\includegraphics[width=3.2in]{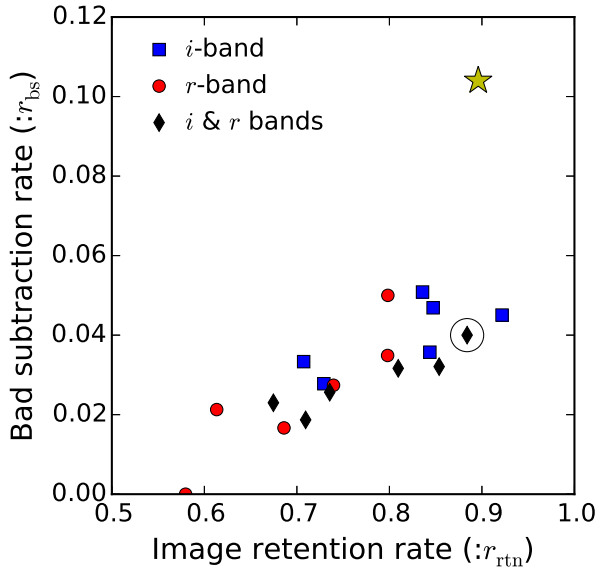}
\caption{\small 
Yellow star: the performance of the system before filtering process (raw data).
Blue squares: the wrapper performance estimations based on the $i$-band data as test set, 
while the $r$-band data was used as the training set (a 2-fold cross validation).
Red circles: the wrapper performance estimations estimated based on the 
$r$-band data as test set, while the $i$-band data was used as the training set. 
Black diamonds: the wrapper performance estimations estimated based on both $i$ and $r$ bands. 
The circled black diamond specifies our desired system efficiency.
{\it The wrapper delivers a robust performance independent of the input.}}
\label{fig:roc}
\end{figure}

The expected performance of the image subtraction wrapper
as the C.I. is varied can be visualized (Figure \ref{fig:roc})
by plotting the {\it Bad} subtraction rate versus image retention rate.
We define in Equation \ref{eq:sen} the image retention rate, $r_{\rm rtn}$, 
of a filter as the ratio of number of accepted \textit{Good} subtracted frames, 
$G_i$, over the total number of input frames.
Similarly, the {\it Bad} subtraction rate, $r_{\rm sb}$, is defined as the ratio of 
the number of accepted \textit{Bad} frames, $B_i$, over the total number 
accepted frames, $B_i + G_i$ (Eq. \ref{eq:far}). Subscripts $i$ and $o$ 
represent data inside and outside of the $[a_k,b_k]$ interval, respectively.

\begin{gather}
i \in [a_k,b_k] \;\;\;\;\;\;\; o \not\in [a_k,b_k]  \notag \\
\mathrm{Image \; retention \; rate}: r_{\rm rtn} = \left( \frac{G_i}{\sum frames} \right) \label{eq:sen}  \\
Bad \mathrm{\; subtraction \; rate}: r_{\rm sb} = \left(\frac{B_i}{B_i + G_i}\right) \label{eq:far}
\end{gather}

Blue squares in Figure \ref{fig:roc} represent the data points based on the $i$-band 
data as the test set and red circles show the data points based on the $r$-band data as the test set. 
Black diamonds correspond to use of both $r$ and $i$ bands.
The tight clustering of these three curves in Figure \ref{fig:roc} illustrates that
the filtering system is quite robust with respect to which data are used to train 
and test and can achieve high image retention rates ($\sim 90$\%) with low 
bad subtraction rates ($\sim 4$\%).
 For comparison purposes, the efficiency of the system prior to the filtering process, raw data, 
 is also displayed on the Figure \ref{fig:roc} with a yellow star.
 We adopt the circled black diamond in Figure \ref{fig:roc} as our final bad subtraction filter.
To reach this performance, the wrapper sets the following constraints for each of the 
{\it Par$_1$}, {\it Par$_2$}, and {\it Par$_3$} values
as [0.64, 1.45], [0.53, 2.5], and [0.55, 2.17], respectively.

\begin{figure}[!ht]
\centering
\includegraphics[width=3.2in]{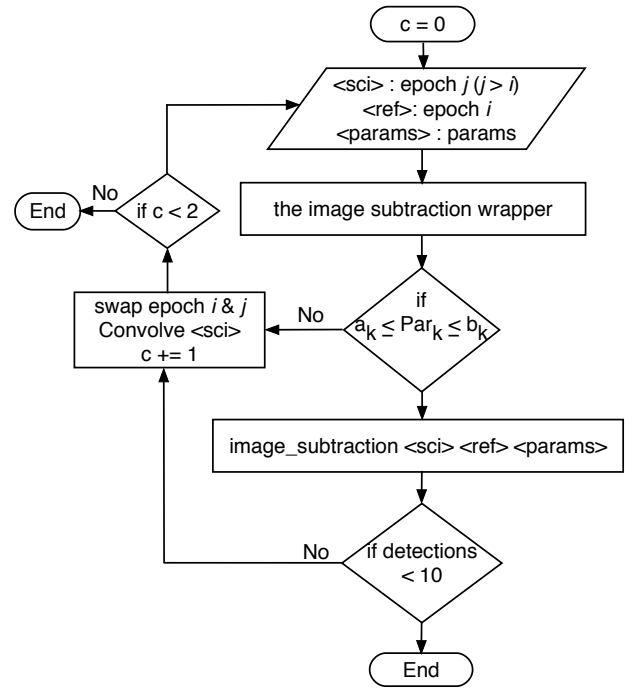}
\caption{\small 
{\it Flowchart}. The procedure to determine how best to subtract two images to yield a reliable result (detections $< 10$).
}
\label{fig:flowchart}
\end{figure}

For each of the imaged fields on two different epochs in our data set, 
we may now identify the optimal subtraction approach prior to performing the subtraction.
This procedure is described using a {\it flowchart} presented in the Figure \ref{fig:flowchart}.
The detailed steps in the {\it flowchart} are performed for each of the imaged fields in the $i$ and $r$ bands, separately. 
We note that, as the last step, a visual inspection is conducted to remove frames with spurious features like satellite trails.

\section{Results \& Discussion}
\label{sec:RD}

RATIR performed rapid-response followup to the second GW trigger released to 
the EM partners by the LIGO team during the \textit{O1} operating run. 
The observing time constraints allowed us to image 26 galaxies ($D < 10$ Mpc; Table 1).
The candidate fields were followed for two more consecutive nights.
We imaged each of the 26 fields in both the $i$ and $r$ bands and 
performed our modified image subtraction routine to search for any possible transients. 

\begin{figure}[!ht]
\centering
\includegraphics[width=3.2in]{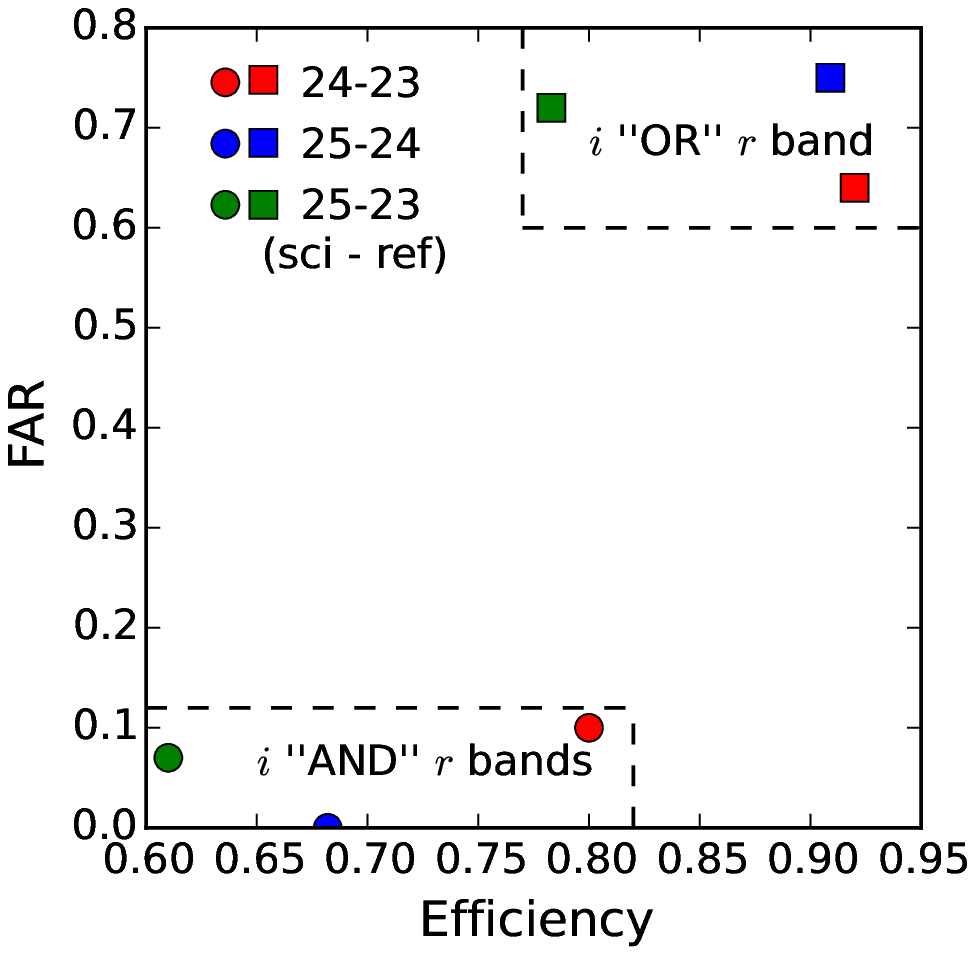}
\caption{\small 
FAR and efficiency of the designed EM counterparts to GW events pipeline.
Candidate galaxies are imaged on three subsequent epochs, 2015/10/23, 10/24, and 10/25.
Algorithm throughput represents with circles if we require both $i$ ``AND'' $r$ bands and
with squares if we require either $i$ ``OR'' $r$ band. 
The combination of multiple frames drives the {\it FAR} 
down as $\lessim 0.1^{N_f -1}$ ($N_f$ depicts the number of available filters). 
}
\label{fig:far}
\end{figure}

For the 25 imaged fields on 10/23 and 10/24 nights, we found only 
two fields (out of 20) yielded more than zero detections at the same sky position in the both $i$ and $r$ bands.
From this, we can estimate a false-alarm-rate (FAR) of 2/20 = 0.1 and an efficiency of 20/25 = 0.8.
This is an ``AND'' rule for comparing the $r$ and $i$ bands to find detections.
We can also consider an ``OR'' rule, either $i$ or $r$ band detections.  We find
FAR and efficiency values of 0.64 and 0.92, respectively, in this case.
These results and those including the 10/25 night are shown in the Figure \ref{fig:far}.
The advantage of multi-bands imaging is clear.
\\
Inspections of all the imaged fields on 10/25 epoch demonstrate a higher 
level of background noise --  compared to the two previous epochs --
due to the non-photometric conditions.
This effect impacts primarily the detection efficiency as shown in the Figure \ref{fig:far}.

Since the G19475 trigger was not of astrophysical origin, we did not expect to ascertain 
a real transient associated with the event. Therefore, we are not able to determine a true sensitivity.
To validate the sensitivity of our approach, we analyze a set of supernova images 
captured with RATIR (PI, Ori Fox). The SNe images were taken using both RATIR's optical 
and Infra-red bands.  Studying 5 SNe fields in the {\it irZYJH} bands for two different epochs, 
we verified that all difference frames containing a $10\sigma$ flux excess 
are indeed recovered via our methodology.

Similar to \citet{2015AJ....150..172K}, 
we find the image subtractions efficiency in our proposed pipeline would not degrade 
with increasing galaxy surface brightness at the lower redshifts ($z < 0.5$), 
which is the case in our search. Therefore, the bright-galaxy issue has 
minimal impact on our discovery program here. 

\subsection{End-to-End Sensitivity and False Alarm Rate}

Given RATIR's simultaneous optical/NIR observing capability, an optimal strategy -- which reduces {\it FAR} as much as possible while keeping sensitivity high -- would be searching for either optical or NIR detections (3 or more bands), considering separately {\it riZ} and {\it YJH}.  This can be viewed as a sGRB strategy with optimal depth (using the {\it riZ} filters), combined with a simultaneous NIR ({\it YJH} band) survey to find potential kilonovae.
Extrapolating our results from optical bands ($i$ and $r$), we expect a 3-band survey to exhibit a 
$ F\!A\!R \approx 1\%$ and an image-retention efficiency  $\approx 90\%$, provided that multiple nights are allocated to re-observe targets missed initially.
We note that a simultaneous multicolor approach is typically employed to increase the probability of detection \citep[see e.g.][]{1999AJ....117...68S}; however, our approach is somewhat different in that we utilize the multiple bands to maintain ($10\sigma$) detection capability while greatly minimizing false alarms.  This is a crucial aspect of the RATIR followup engine as it will add the most confidence to justify additional observing requests with larger aperture facilities.
We have rigorously tested the tradeoff between false alarm rate and sensitivity in the design of our implemented methodology.

In order to estimate the total system throughput for potential EM gravitational wave counterparts detected by RATIR, we adopt a similar LIGO (and Virgo) error region fraction value as \citet{2016ApJ...820..136G} for {\it O2} (2017) $\sim100 \, \mathrm{deg^2}$.
If we utilize the GWGC nearby galaxy catalog and select the brightest galaxies representing half of the population light (1 galaxy per $\mathrm{deg^2}$, \citealt{2011CQGra..28h5016W}), 
we can concentrate observations on about 100 galaxies in the LIGO (and Virgo) 3$\sigma$ error region. 
Given that not all the candidate galaxies in our list will be visible at the RATIR observatory site, we expect the candidates list to be reduced by about half.
As mentioned in the Section 1, RATIR reaches its design 10$\sigma$ limiting magnitude in about 10 minutes, and the candidate sources are expected to be brighter than our limiting magnitudes.
Allowing for a conservative overhead of 20\% for slew between galaxy positions, we expect to be able to survey all the galaxies 
in about 10 hrs. In this case of narrower regions, we are able to integrate more deeply and also to check for same variability same night. Our nominal strategy is then to re-observe the field the next night (or the night after in the case of poor weather) to check for variability. 

The brighter galaxy selection of the GWGC maintains a galaxy completeness near 100\% to a distance of 
$\sim$90 Mpc and drops to 85\% at the distance of $\sim$100 Mpc as the upper bound region.
The error region source fraction resulting from the wrapped image subtraction approach reduces the search space region to about 90\% as the image retention rate. Our automated pipeline flags $\sim$1\% of the candidate imaged fields as potential targets for the additional followup. 
Combining the galaxy completeness ($\approx 50$\%) with the RATIR survey completeness ($\approx 45$\%), limited primarily by observability, we estimate a final success rate of $\approx 20$\%.
This is an excitingly large number, given the great scientific impact of an identified EM counterpart to a gravitational wave event.  The possibility of multiple followup campaigns of separate triggers only increases our chances.  
In 2016/2017 RATIR observed additional LIGO fields, and analysis is underway.

As LIGO/Virgo sensitivity increases, we will require a galaxy catalog which covers distances up to $\sim$450 Mpc \citep{2013CQGra..30l3001B} with greater completeness than is currently available.
There are some attempts to combine other galaxy catalogs e.g., the 2MASS Photometric Redshift Catalog 
\citep[2MPZ;][]{2014ApJS..210....9B} with the GWGC, as a short-term plan 
\citep[see also][]{2016ApJ...820..136G,2016yCat.7275....0D} 
and also efforts to constrain the exact GW distance scale \citep{2016ApJ...829L..15S}.
We plan to adopt a more complete galaxy catalog in our pipeline as it becomes available.
We have been actively following LIGO triggers during the {\it O2} run as well as candidate counterparts reported by
other EM followup groups. The results of our observations reported in LVC/GCN circulars (e.g., \citeauthor[][GCN 20485]{2015GCN..20485...2G}). 

Combining a distance and position information from the GW observations provided 
during LIGO/Virgo {\it O2} run with our complied list of nearby galaxies can 
reduce the search space and help to prioritize targets for further followup. 
Since we are more interested in the impact of the galaxy catalog we follow 
a similar strategy as \citet{2016ApJ...820..136G} and assume that the localization subtends a given solid angle, 
which is a shell between two constant radii.
Having a target distance information such as a BBH event at a distance of 350 Mpc 
with an uncertainty of 25-30\% 
\citep[see for example,][]{1994PhRvD..49.2658C,2015ApJ...804..114B}
would reduce the volume, 
and hence number of galaxies, by a factor of 50-60\%. 
This is in agreement with the finding of \citet{2013ApJ...767..124N}. 

RATIR multicolor observations can provide powerful discriminating 
information for candidates found by other facilities.  
In the case of immediate, robotically-triggered observations following a LIGO detection, 
RATIR's narrow field of view ($\lessim 10$ arcmin) dictates a search strategy 
which targets nearby galaxies within the large LIGO error circles.  
These nearby events are, in turn, the most likely to yield decisive CBC associations. 
As a 6-filter, multicolor instrument,  RATIR's simultaneous observations of candidates greatly reduces
the number of false alarms, while also providing spectral information 
and additional time-sampling \citep[see e.g.,][]{2014ApJ...787...90G,2014AJ....148....2L,2015ApJ...811...93G}, important for afterglow studies. 
Moreover, with the recent clue that some sGRBs may be associated
with very red ``kilonova'' events (\citealt{2013ApJ...775...18B,2013Natur.500..547T}, see also \citealt{2016NatCo...712898J}), 
NIR observations may be essential to afterglow detection.

\acknowledgments
We thank the RATIR project team and the staff of the Observatorio Astron\'omico 
Nacional on Sierra San Pedro M\'artir, and acknowledge the contribution of 
Leonid Georgiev to its development.  RATIR is a collaboration between 
the University of California, the Universidad Nacional Auton\'oma de M\'exico, 
NASA Goddard Space Flight Center, and Arizona State University, 
benefiting from the loan of an H2RG detector and hardware and software 
support from Teledyne Scientific and Imaging. RATIR, the automation 
of the Harold L. Johnson Telescope of the Observatorio Astron\'omico 
Nacional on Sierra San Pedro M\'artir, and the operation of both are funded 
through NASA grants NNX09AH71G, NNX09AT02G, NNX10AI27G, and 
NNX12AE66G, CONACyT grants INFR-2009-01-122785 and CB-2008-101958, 
UNAM PAPIIT grant IN100317, and UC MEXUS-CONACyT grant CN 09-283.
We also thank our referee for insightful comments that improved the manuscript.


\vfill
\eject
\begin{deluxetable*}{rrr|rr|rr|rr}
    \tablecolumns{9}
    \tablewidth{\textwidth}
    \tablecaption{\label{table:medians}Observations Log}
    \tablehead{
        \colhead{} &
        \colhead{} &
        \colhead{} &
        \multicolumn{2}{c}{20151023} &
        \multicolumn{2}{c}{20151024} &
        \multicolumn{2}{c}{20151025} 
        \\
        \colhead{\#} &
        \colhead{RA (J2000)} &
        \colhead{DEC (J2000)} &
        \multicolumn{2}{c}{total exp. (sec)} &
        \multicolumn{2}{c}{total exp. (sec)} &
        \multicolumn{2}{c}{total exp. (sec)} 
        \\
        \cline{4-5}
        \cline{6-7}
        \cline{8-9}
        \colhead{ } &
        \colhead{ } &
        \colhead{ } &
        \colhead{$r$-band} &
        \colhead{$i$-band} &
        \colhead{$r$-band} &
        \colhead{$i$-band} &
        \colhead{$r$-band} &
        \colhead{$i$-band} 
    }
    \startdata
    1 & 0.493792 & -15.461389 & 480.0 & 480.0 & 560.0 & 560.0 & 720.0 & 720.0\\[4pt]
    2 & 2.485667 & -24.963111 & 480.0 & 480.0 & 480.0 & 480.0 & 640.0 & 720.0\\[4pt]
    3 & 3.516292 & -23.182111 & 240.0 & 240.0 & 320.0 & 400.0 & 720.0 & 640.0\\[4pt]
    4 & 3.855459 & -21.444805 & 480.0 & 480.0 & 240.0 & 240.0 & 480.0 & 400.0\\[4pt]
    5 & 4.797917 & -22.668389 & 480.0 & 480.0 & 240.0 & 320.0 & 480.0 & 480.0\\[4pt]
    6 & 8.70375 & 7.450389 & 720.0 & 720.0 & 960.0 & 960.0 & -- & -- \\[4pt]
    7 & 10.765958 & -22.246806 & 240.0 & 240.0 & 480.0 & 480.0 & 640.0 & 640.0\\[4pt]
    8 & 11.785666 & -20.760389 & 480.0 & 480.0 & 400.0 & 240.0 & 720.0 & 560.0\\[4pt]
    9 & 11.888083 & -25.288805 & 240.0 & 240.0 & 240.0 & 240.0 & 480.0 & 480.0\\[4pt]
    10 & 12.338083 & -18.075889 & 480.0 & 480.0 & 240.0 & 400.0 & 720.0 & 720.0\\[4pt]
    11 & 12.457792 & -21.012194 & 720.0 & 720.0 & 480.0 & 480.0 & 800.0 & 800.0\\[4pt]
    12 & 12.60225 & -19.906194 & 720.0 & 720.0 & 480.0 & 480.0 & 960.0 & 960.0\\[4pt]
    13 & 12.800083 & 12.024611 & 1200.0 & 1200.0 & 880.0 & 800.0 & 1200.0 & 880.0\\[4pt]
    14 & 16.225667 & 2.133305 & 1200.0 & 1200.0 & 960.0 & 960.0 & 1200.0 & 1200.0\\[4pt]
    15 & 16.943708 & 1.0635 & 1200.0 & 1200.0 & 960.0 & 960.0 & 1200.0 & 1120.0\\[4pt]
    16 & 20.3295 & 12.411694 & 1440.0 & 1440.0 & 880.0 & 880.0 & 1200.0 & 1200.0\\[4pt]
    17 & 22.828792 & 7.787694 & 1440.0 & 1440.0 & 960.0 & 960.0 & 1200.0 & 1200.0\\[4pt]
    18 & 352.150792 & 14.743 & 480.0 & 480.0 & 480.0 & 480.0 & 720.0 & 720.0\\[4pt]
    19 & 7.464167 & -16.165111 & 480.0 & 480.0 & 480.0 & 480.0 & 640.0 & 640.0\\[4pt]
    20 & 6.54525 & -11.053889 & 480.0 & 480.0 & 640.0 & 640.0 & -- & -- \\[4pt]
    21 & 5.965667 & -24.705111 & 480.0 & 480.0 & 400.0 & 400.0 & 480.0 & 480.0\\[4pt]
    22 & 2.30025 & -26.161111 & 240.0 & 240.0 & 320.0 & 320.0 & 480.0 & 400.0\\[4pt]
    23 & 5.173792 & 8.615389 & 720.0 & 720.0 & 960.0 & 720.0 & -- & -- \\[4pt]
    24 & 13.197917 & -26.59 & 480.0 & 480.0 & 320.0 & 320.0 & 560.0 & 560.0\\[4pt]
    25 & 346.6845 & 12.771889 & 480.0 & 480.0 & 480.0 & 480.0 & 720.0 & 720.0\\[4pt]
    26 & 347.1105 & -15.611389 & 240.0 & 240.0 & -- & -- & 480.0 & 480.0\\[2pt]
    \enddata
\end{deluxetable*}

\end{document}